\documentclass[epp]{agutex}

\usepackage{multirow,color}
\usepackage{bbding}
\usepackage{amssymb}
\usepackage{ulem}
\usepackage{verbatim}
\usepackage{url}

\usepackage{graphicx}
\usepackage[displaymath]{lineno}

\def\gsim{\;\lower4pt\hbox{${\buildrel\displaystyle >\over\sim}$}\;}
\def\lsim{\;\lower4pt\hbox{${\buildrel\displaystyle <\over\sim}$}\;}
\def\grls{\;\lower4pt\hbox{${\buildrel\displaystyle >\over <}$}\;}

\newcommand{\del}[1]{{\color[rgb]{0,0,1} \sout{#1}}}
\renewcommand{\del}[1]{{}}

\newcommand{\ve}[1]{\mathbf{#1}}

\volume{0}
\authorrunninghead{WANG ET AL.}
\titlerunninghead{Locating DAM Source}

\begin{document}

\title{Locating the Source Field Lines of Jovian Decametric Radio Emissions}

\author{Yuming Wang,$^{1,2,3,*}$ Xianzhe Jia,$^{4}$ Chuanbing Wang,$^{1,2}$ Shui Wang,$^{1,2}$ and V. Krupar$^{5,6,7}$}

\affil{$^1$ CAS Key Laboratory of Geospace Environment, School of Earth and Space Sciences, University of Science and Technology of China, Hefei 230026, China}

\affil{$^2$ CAS Center for Excellence in Comparative Planetology, Hefei 230026, China}

\affil{$^3$ Mengcheng National Geophysical Observatory, School of Earth and Space Sciences, University of Science and Technology of China, Hefei 230026, China}

\affil{$^4$ Department of Climate and Space Sciences and Engineering, University of Michigan, Ann Arbor, MI 48109-2143, USA}

\affil{$^5$ Universities Space Research Association, Columbia, Maryland, USA}

\affil{$^6$ NASA Goddard Space Flight Center, Greenbelt, Maryland, USA}

\affil{$^{7}$ Department of Space Physics, Institute of Atmospheric Physics, The Czech Academy of Sciences, Prague, Czech Republic}

\affil{$^*$ Corresponding author, Email: ymwang@ustc.edu.cn}

\begin{abstract}
Decametric (DAM) radio emissions are one of the main windows through which one can reveal and understand the Jovian
magnetospheric dynamics and its interaction with the moons. DAMs are generated by energetic electrons through
cyclotron-maser instability. For Io (the most active moon)
related DAMs, the energetic electrons are sourced from Io volcanic activities, and quickly trapped by
neighboring Jovian magnetic field. To properly interpret the physical
processes behind DAMs, it is important to precisely locate the source field lines from which DAMs are emitted.
Following the work by~\citet{Hess_etal_2008, Hess_etal_2010}, we develop a method to locate the source region
as well as the associated field lines
for any given DAM emission recorded in a radio dynamic spectrum by, e.g., Wind/WAVES or STEREO/WAVES.
The field lines are calculated by the state-of-art analytical model, called JRM09~\citep{Connerney_etal_2018}.
By using this method, we may also derive the emission cone angle and the energy of associated electrons.
If multiple radio instruments at different perspectives saw the same DAM event, the evolution
of its source region and associated field lines is able to be revealed. We apply the method
to an Io-DAM event, and find that the method is valid and reliable. Some physical processes behind the DAM event
are also discussed.
\end{abstract}

\begin{article}

\section{Introduction}

Jupiter is the largest planet in our solar system with fastest rotation and strongest magnetic field.
Its surface magnetic field is as strong as tens Gauss, which leads to the most intense planetary decametric (DAM)
radio emissions~\citep[e.g.,][and references therein]{Zarka_1998}. Like the auroral kilometric radiations (AKRs)
at the Earth, Jovian DAM emissions are caused by high energetic electrons through Cyclotron-Maser (CM)
instability~\citep[e.g.,][]{Wu_Lee_1979, Lecacheux_1988, Dulk_etal_1992, Queinnec_Zarka_1998, Treumann_2006, Hess_etal_2008}.
The energy of electrons emitting DAMs are believed to be generally above 0.5 keV~\citep{Waite_etal_1988, Zarka_etal_1996,
Hess_etal_2008}.
Since Jupiter's magnetosphere holds cold and strongly magnetized plasmas with the electron plasma
frequency much smaller than electron cyclotron frequency
(i.e., $f_{pe}/f_{ce}<0.1-0.2$)~\citep[e.g.,][]{Connerney_1992, Bagenal_1994}, Jovian DAMs beam along the narrow
walls~\citep[e.g.,][]{Kaiser_etal_2000, Panchenko_Rucker_2016}
of a widely-opened hollow cone according to the basic properties of CM
emissions~\citep[e.g.,][]{Queinnec_Zarka_1998, Hess_etal_2008, Hess_etal_2014, Lamy_etal_2008, Lamy_etal_2013}.
This process leaves DAM emissions an `arc' shaped drift pattern in the radio dynamic spectrum, ranging roughly
from one to several tens MHz depending on the magnitude of the magnetic field and the energy of the associated electrons.

On the other hand, unlike AKRs which are mainly driven by the external forcing, i.e., the solar wind, the Jovian DAMs, especially
the Io-related DAMs occurring in the inner magnetosphere, are mostly driven by
internal processes due to the strong magnetic field~\citep[e.g.,][]{Hill_etal_1983, Zarka_1998,
Cowley_Bunce_2001, Kivelson_Southwood_2005}. Io, with more than 400 volcanoes, is the most active moon of Jupiter,
supplying large amount of plasmas into magnetosphere~\citep{Schneider_Bagenal_2007}.
These loaded plasmas are picked up by and corotate with the Jovian magnetosphere, and drive many dynamic processes
there including DAMs. Thus, DAM emissions are of great interest as they carry important information of the Jovian
magnetospheric dynamics, and can be detected by remote radio instruments on ground and/or on spacecraft.

One of the key information of DAMs is the source location as well as the associated field lines from which they are emitted.
According to the `arc'-shaped drift pattern of DAMs on radio dynamic spectra, we can know if they come from
the vicinity of west limb (look like `a front bracket') or east limb (`a rear bracket') of Jupiter. However,
such a determination is too rough to reveal detailed processes behind the observed features. For example, the plasma in the
flux tube linked to Io is believed to be the source of Io-DAMs. However, the strongest emission of Io-DAMs typically
shifts toward smaller central meridian longitudes (CML) in System III, i.e., moving ahead of Io in the direction
of Io orbiting around Jupiter~\citep[e.g.,][]{Zarka_1998, Imai_etal_2002, Hess_etal_2010}. Such a shift, or called lead angle
between the active field line and Io flux tube, can be explained by
the presence of Alfv\'en wings arising from Io's interaction with the sub-Alfv\'enic flow of the Jovian
corotating plasma~\citep[e.g.,][]{Hill_etal_1983, Jacobsen_etal_2007, Bonfond_etal_2008, Bonfond_etal_2009}.

To locate the DAM radio source, \citet{Imai_etal_1997, Imai_etal_2002} developed a model based on
the observed modulation lanes in the high frequency range of DAM spectra~\citep[e.g.,][]{Riihimaa_1968, Riihimaa_1978}.
In that model, the source is assumed to be a linear and thin structure along the Jupiter's tilted dipole magnetic field
and corotate with Io. The emission cone angle and the associated longitude of the emission can be derived from this model.
Later, \citet{Hess_etal_2008, Hess_etal_2010} established a relation between the velocity of energetic electrons
and the emission cone angle based on the lose-cone distribution, and then developed a general simulation code called
`ExPRES' (Exoplanetary and Planetary Radio Emission Simulator). In this method, the lead angle between the
active field line and Io in longitude and the energy of the associated electrons are two free parameters obtained by
fitting observed DAM arcs.
Here inspired by \citet{Hess_etal_2008, Hess_etal_2010}, we propose a similar method
to determine the source region of a DAM, in which a state-of-art Jovian
magnetic field model~\citep{Connerney_etal_2018} is used. As will be shown below, the core function is the same
as that of \citet{Hess_etal_2008, Hess_etal_2010}, and the input is also just the observed
DAM arcs in a broad frequency band, but the procedure is different and therefore results in similar
but not exactly same output. Our method, for example, does not introduce the lead angle between the active field lines
and Io as a free parameter when locating DAM source, though we may derive it after all the source field lines have been identified.
With observations from multiple spacecraft at different perspectives,
the evolution of the DAM source, including, e.g., the apparent rotation speed of the source, the range of the active
field lines and the energy of associated electrons, can be revealed.
The details of the method are presented in the next section.
In Sections~\ref{sec:case} and~\ref{sec:analysis}, we will verify the method by applying it to an Io-DAM and show some
interesting processes behind the Io-DAM. A summary and discussion is then given in Section~\ref{sec:summary}.

\section{Method}~\label{sec:method}

\begin{figure*}[tbh]
\begin{center}
\includegraphics[height=0.6\vsize]{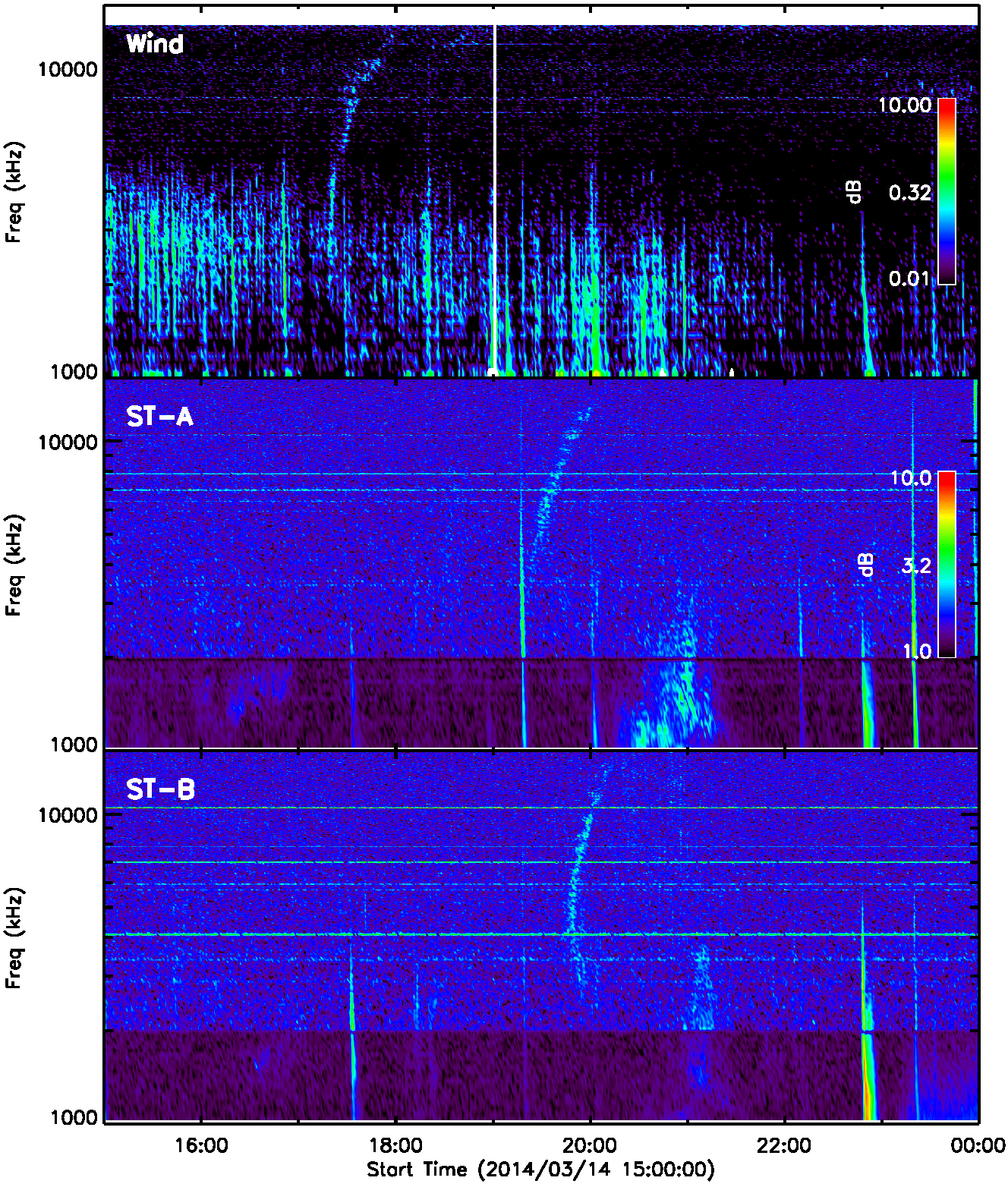}
	\caption{From top to bottom, the panels show the radio dynamic spectra from Wind/WAVES, ST-A/WAVES and ST-B/WAVES,
respectively.}\label{fig:rds}
\end{center}
\end{figure*}

The basic idea of our locating method is that a DAM emission is not isotropic and the frequency depends on the
local magnetic field strength owing to the fact that DAMs are fundamental X mode through the CM
instability~\citep[e.g.,][]{Lecacheux_1988, Zarka_1998}.
From a radio dynamic spectrum, we can read the frequency-time drift pattern of a DAM as shown in Figure~\ref{fig:rds}.
First, we may immediately know that the DAM comes from eastern/western hemisphere if the drift pattern looks like a
rear/front bracket~\citep[e.g.,][]{Hess_etal_2014}. Second, the recorded frequency can be used to derive the magnetic
field strength in the source region where the DAM is emitted according to the electron cyclotron frequency formula
\begin{eqnarray}
B=0.358f
\end{eqnarray}
in which $B$ is in units of Gauss and $f$ in units of MHz. It should be noted that the observed frequency $f$ is larger than
but very close to the local electron cyclotron frequency $f_{ce}$ based on theory, the magnetic field $B$ is therefore
slightly overestimated, which is ignored in our method. After converting the observed frequency to magnetic field strength,
we may search the Jupiter's eastern or western magnetosphere for all the places matching the requirement of $B$.

\begin{figure}[tbh]
\begin{center}
\includegraphics[width=0.9\hsize]{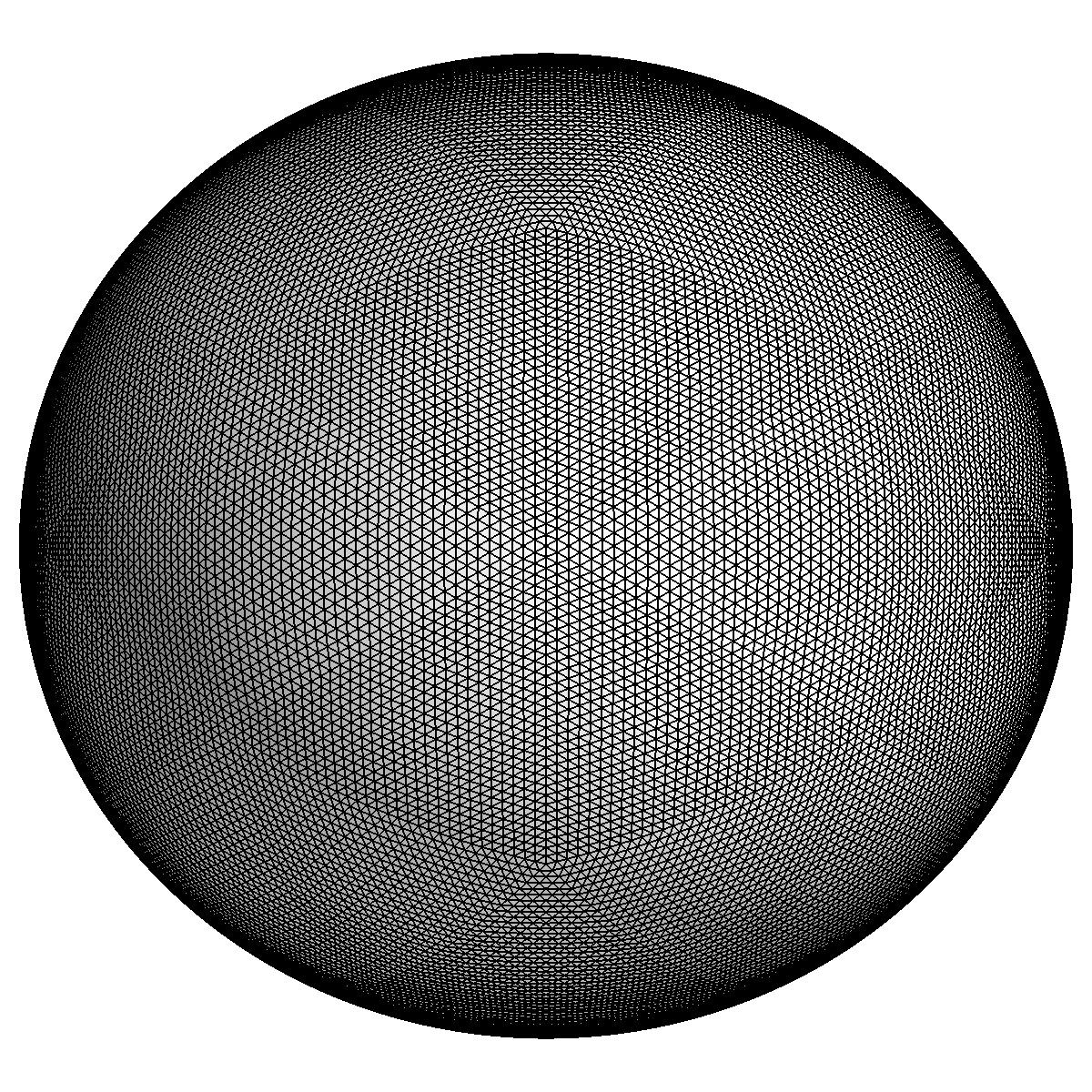}
	\caption{The flattened geodesic polyhedron with 40962 vertices (see main text for details). Since Jupiter is an ellipsoid with the equatorial radius
of $1 \rm{R_J}=71492$ km and polar radius of $0.935\rm{R_J}$, the polyhedron is flattened by a factor of $1/15.4$.}\label{fig:icosphere}
\end{center}
\end{figure}

The Jupiter's magnetic field is calculated by using the newest model based on the observations from Juno's first nine
orbits~\citep{Bolton_etal_2017}.
The model consists of two components. One is the potential field originating from interior of Jupiter, of which the
spherical harmonic coefficients extending to the order of 10 were updated by \cite{Connerney_etal_2018}, and referred as
`JRM09' model field (Juno Reference Model through Perijove 9). The other is the
external field due to the presence of the magnetodisc (or the current sheet) from 5 to several tens of Jovian radii~\citep{Connerney_etal_1981}.
Here we use the typical values of the parameters of the magnetodisc model, i.e., the total current related coefficient $\mu_0I_0/2=225$,
the inner radius of the disc $r_a=5\rm{R_J}$, the outer radius $r_b=50\rm{R_J}$, and semi-thickness $D=2.5\rm{R_J}$.
A more accurate formula to calculate the external field was given by \cite{Giampieri_Dougherty_2004}, and is adopted here.
By using the JRM09 plus magnetodisc model, we trace magnetic field lines from the $1/15.4$ flattened surface of one Jovian radius ($\rm{R_J}$) to get the
$\ve B$ vectors including the strength and direction in the Jovian magnetosphere. The spatial step to trace the field lines
is ranged between 0.01 and 0.5 $\rm{R_J}$ depending on the curvature of the local field line. To
avoid possible biases from a non-uniform distribution of footpoints of the field lines, we use a geodesic
polyhedron\footnote{see wikipedia \url{https://en.wikipedia.org/wiki/Geodesic_polyhedron} for details} with 40962 vertices (see
Fig.\ref{fig:icosphere}), from which we trace all the field lines. The number of footpoints makes the angle between neighboring points
being about $1^\circ$, which meets the angular resolution required from the thickness of the wall of the emission cone as discussed below.

A DAM is emitted along the narrow wall of a hollow cone, meaning that an observer can only receive the emission from a certain direction.
The emission cone angle, $\alpha$, generally varies between
$60^\circ-90^\circ$~\citep[e.g.,][]{Queinnec_Zarka_1998, Hess_etal_2008, Hess_etal_2014, Lamy_etal_2008, Lamy_etal_2013} with
the wall thickness less than $2^\circ$~\citep[e.g.,][]{Kaiser_etal_2000, Panchenko_Rucker_2016}.
This is another constraint for our method to locate the DAM source. For each candidate point selected based on the magnetic
field strength, we know the direction of the point-observer line, and therefore the angle, $\alpha_p$, between the line and the
local field line where the point locates. These candidate points are then further filtered by checking if the associated $\alpha_p$
falls in the range of expected $\alpha$ with the thickness of $2^\circ$.

The range of the emission angle $\alpha$ is too wide to make precise filtering. Thus, we follow \citet{Hess_etal_2008,
Hess_etal_2010} works, use the formula below to constrain the value of $\alpha$
\begin{eqnarray}
\alpha=\arccos\left[\frac{v}{c}\left(1-\frac{f_{ce}}{f_{ce,max}}\right)^{-1/2}\right] \label{eq:ea}
\end{eqnarray}
in which $v$ is the speed of the energetic electrons, $c$ is the light speed and $f_{ce,max}$ is the maximum
of the electron cyclotron frequency that the electrons can reach along the active field line, i.e., the high cut-off,
typically choosing the value at the top of ionosphere.
For each field line on which there are some candidate points,
we know the values of $\alpha_p$ of all these candidate points and the associated magnetic field strength, i.e., the frequency $f$,
and can fit the data pair of $\alpha_p$ and $f$ with Equation~\ref{eq:ea} by assuming that the energy, i.e., $v/c$, of the electrons
generating the DAM emission keeps unchanged when moving along this field line. Based on the previous studies mentioned before,
our procedure limits the range of $\alpha$ to $50^\circ-90^\circ$ and the energy of electrons to above $0.2$ keV (or the speed $v>0.05c$).
If the fitting can reach a converged value of $v/c$, say its standard deviation is less than $0.01$ in our procedure, the candidate points
as well as the associated field line are finally selected.

It should be noted that, for one DAM arc in the radio dynamic spectrum, the method may find
a set of field lines matching the above criteria: (1) the strength of the field line covers the
corresponding frequency swept by the DAM arc, and (2) the emission cone angles and corresponding frequencies of
all the candidate points on each field line can be well fitted by Equation~\ref{eq:ea} with a single value of electron velocity.
On each of the selected field lines, the obtained electron velocity may be different. But we think that the electrons on
these identified field lines could all contribute to the observed DAM emission, and therefore could be all real sources.

The above procedure is based on the observed radio dynamic spectrum from one spacecraft. Through it, we may locate the
source region of a DAM emission with the following parameters derived: (1)
the position of the source in three-dimensional space, (2) the information of the associated field lines including the longitudes
and latitudes of their footprints, (3) the emission cone angle and (4) the energy or speed of the associated electrons.
If there are multiple spacecraft at different perspectives, e.g., Wind, STEREO-A and B (ST-A and ST-B for short),
we may further obtain the changes of these derived parameters, reflecting the evolution of the emission source.

Besides, one more constraint is the polarization of emission. As DAMs are the result of CM instability, the
right-handed polarized DAMs originate from northern hemisphere and the left-handed polarized DAMs from southern
hemisphere~\citep[e.g.,][]{Zarka_1998, Treumann_2006}. However, not all of the radio observations have the information
of polarization. For those having no polarization measurements, we either try both hemispheres or pre-determine the
hemisphere based on other information, e.g., the tilt angle of the magnetic pole respective to the observer. If the north pole
tilts toward the observer, we tend to believe that the emission comes from the northern hemisphere.

\section{An Io-DAM on 2014 March 14}~\label{sec:case}

\begin{figure}[tbh]
\begin{center}
\includegraphics[width=\hsize]{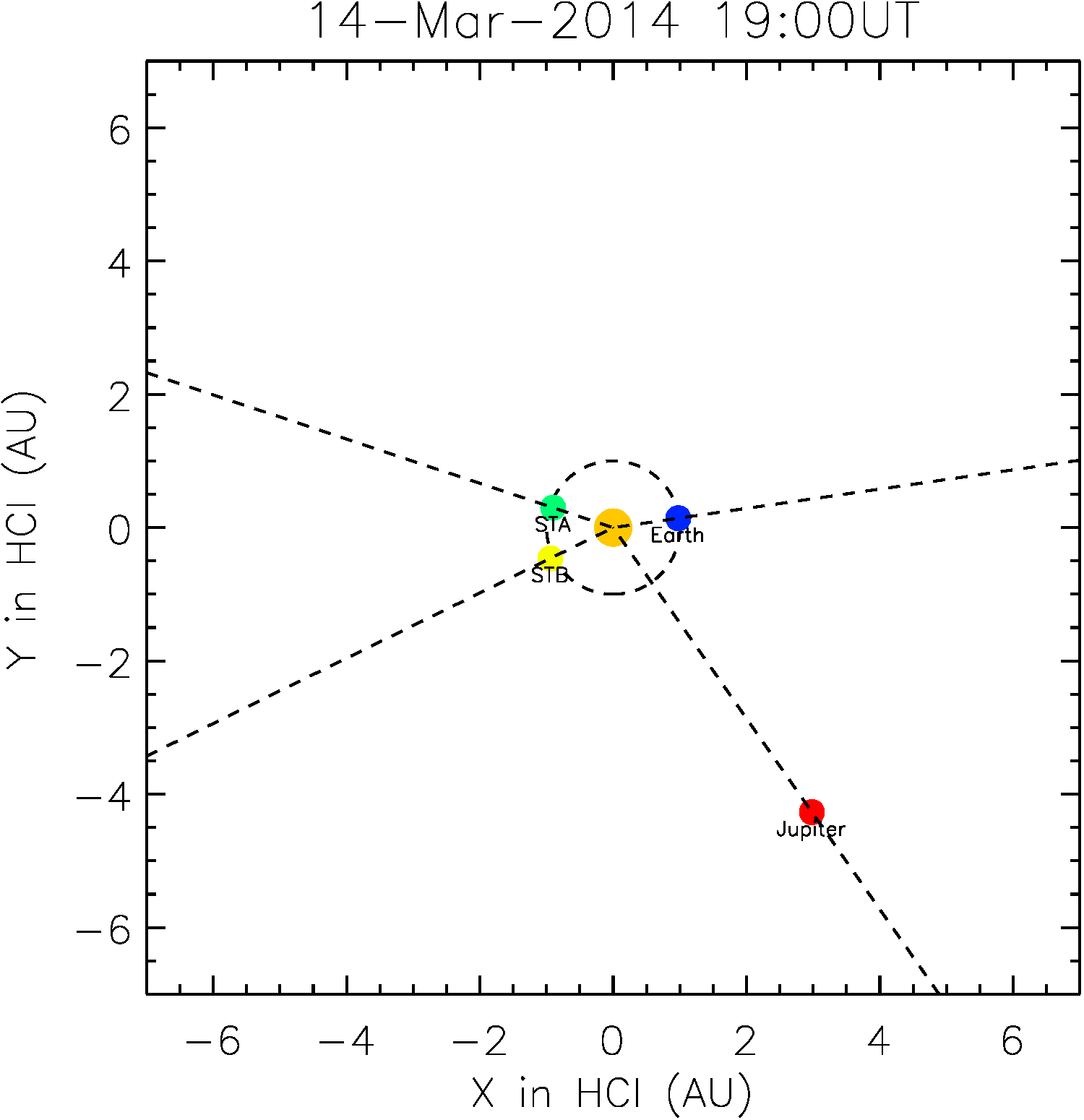}
	\caption{Positions of the Earth, ST-A, ST-B and Jupiter in the heliocentric inertial coordinate (HCI) system.
The big yellow dot indicates the Sun.}\label{fig:pos}
\end{center}
\end{figure}

To verify the method, we apply it to an Io-DAM event. For an Io-DAM, we know the position of Io and therefore roughly
know where its related DAM should be emitted. The Io-DAM observed by Wind/WAVES~\citep{Bougeret_etal_1995} and ST-A and
B/WAVES~\citep{Bougeret_etal_2008} on 2014 March 14
(see Fig.\ref{fig:rds}) is selected for the test. In the radio dynamic spectra, we can see clearly an intense radio
signature drifting from about $5$ to $16$ MHz within about $25$ minutes. The Wind/WAVES first received the DAM emission at
about 17:22 UT with a long-lasting radio noisy burst appearing below $5$ MHz. The ST-A/WAVES received the
DAM emission about two hours later at about 19:24 UT. The ST-B/WAVES received the signal at about 19:44 UT.
The time shifts of the DAM signature on the three spectra are due to the configuration of the three spacecraft relative
to Jupiter as shown in Figure~\ref{fig:pos}. The emission cone of the DAM first swept the Earth, and then ST-A and ST-B.
The time difference due to the different travel distance from Jupiter to observers of the light among the Earth, ST-A and B
is several minutes, which is much shorter compared to the observed time shifts among the observers. After correcting for
the difference due to the light travel time, we deduce that the source of DAM rotated at a speed much slower than Jupiter
itself, but was very close to the rotation speed of Io. Thus, it is Io-related.

\begin{figure}[tbh]
\begin{center}
\includegraphics[width=\hsize]{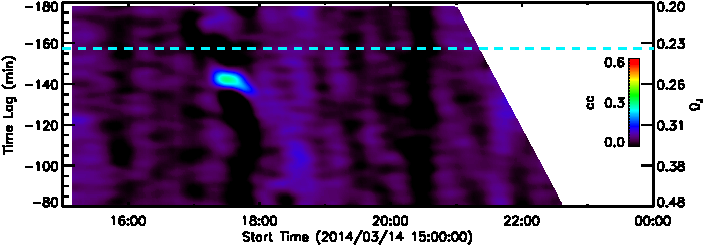}
	\caption{A 2D distribution of the correlation coefficients made between the spectra from Wind/WAVES and
ST-B/WAVES. The horizontal axis gives the time at Wind, and the vertical axis on the left indicates the time lag
of ST-B. For each time lag, we calculate the corresponding rotation speed in units of self-rotational angular speed
of Jupiter, $\Omega_J$, scaled by the vertical axis on the right. The cyan line denotes the rotation speed of Io.
See main text for more details.}\label{fig:corr}
\end{center}
\end{figure}

A more accurate estimation of the rotation speed of the DAM source is made by calculating the
two-dimensional (2D) cross-correlation coefficient (cc) between the radio dynamic spectra from two spacecraft,
say Wind and ST-B. We set a $21$-minute wide and $5$ MHz high box locating between $5$ and $10$
MHz, where the DAM signature is the most significant in both spectra (see Fig.\ref{fig:rds}), and run
the box through both radio dynamic spectra with a certain time shift (or the time lag of the
box in one spectrum compared to the other) to calculate the cc value.
The width of the box is set to be $21$ minutes because it can roughly cover the track of the DAM in the spectra.
For a given time lag, we may get a time sequence of the cc values. Then we adjust the time lag to obtain a 2D cc
distribution as shown in Figure~\ref{fig:corr}, in which the horizontal axis is the time corresponding to the time
recorded by Wind/WAVES and the vertical axis on the left gives the time lag. A negative value of the time lag means
that the running box in the ST-B's spectrum lags behind that in the Wind's spectrum.

One can find one and only one significantly high cc region around the time when the DAM appeared in the Wind's spectrum.
The time lag of the high cc region is about $-142$ minutes, including the time difference of about 5 minutes due to the
light travel. Since the angular separation of the Wind and ST-B with
respective to Jupiter is about $21.5^\circ$, we may derive that the rotation speed of the DAM source, $\Omega_{cc}$, is about
$0.255\Omega_J$, in which $\Omega_J\approx0.6^\circ/$min is the self-rotational angular speed of Jupiter, as scaled by the
vertical axis on the right-hand side. The rotation speed of Io, $\Omega_{Io}$, is about $0.23\Omega_J$, which is marked by the
cyan dashed line. According to the statistical study of UV observations of Jupiter from Habble Space
Telescope~\citep{Grodent_etal_2008, Bonfond_etal_2009},
the rotation speed of the Io's footprint (IFP) magnetically mapping on the northern ionosphere,
$\Omega_{IFP}$, is about $0.55\Omega_J$ during the event.
The estimated rotation speed of the DAM source is bounded between the two speeds, suggesting
that the DAM emission was Io related.

We then apply our method to this event to identify the source region from which the DAM was emitted.
From the drift shape of this DAM emission, we know it came from the western
hemisphere. We do not have the polarization measurements of the event, but on 2014 March 14 the northern magnetic pole
was tilted toward the spacecraft. Thus, we believe that the DAM emission should come from the northern hemisphere. Previous
studies also suggested that DAMs are more likely to come from the northern hemisphere under such configurations~\citep[e.g.,][]{Carr_etal_1983}.
We use the drifting tracks above $5$ MHz for this event, as they are most clear in the radio dynamic spectra. These tracks have a
width of about $10$ minutes at a given frequency and span over about $25$ minutes from $5$ to $16$ MHz. The value of $f_{ce,max}$ is set
to be the value at the footprint of each field line on the $1/15.4$ flattened surface of one $\rm{R_J}$, as an approximation of the
frequency at the top of ionosphere.
Then we input all these information
into our method, and search all the field lines starting from the vertices (see Fig.\ref{fig:icosphere}) on the western and
northern hemisphere.

\begin{figure*}[tb]
\begin{center}
\includegraphics[width=0.7\hsize]{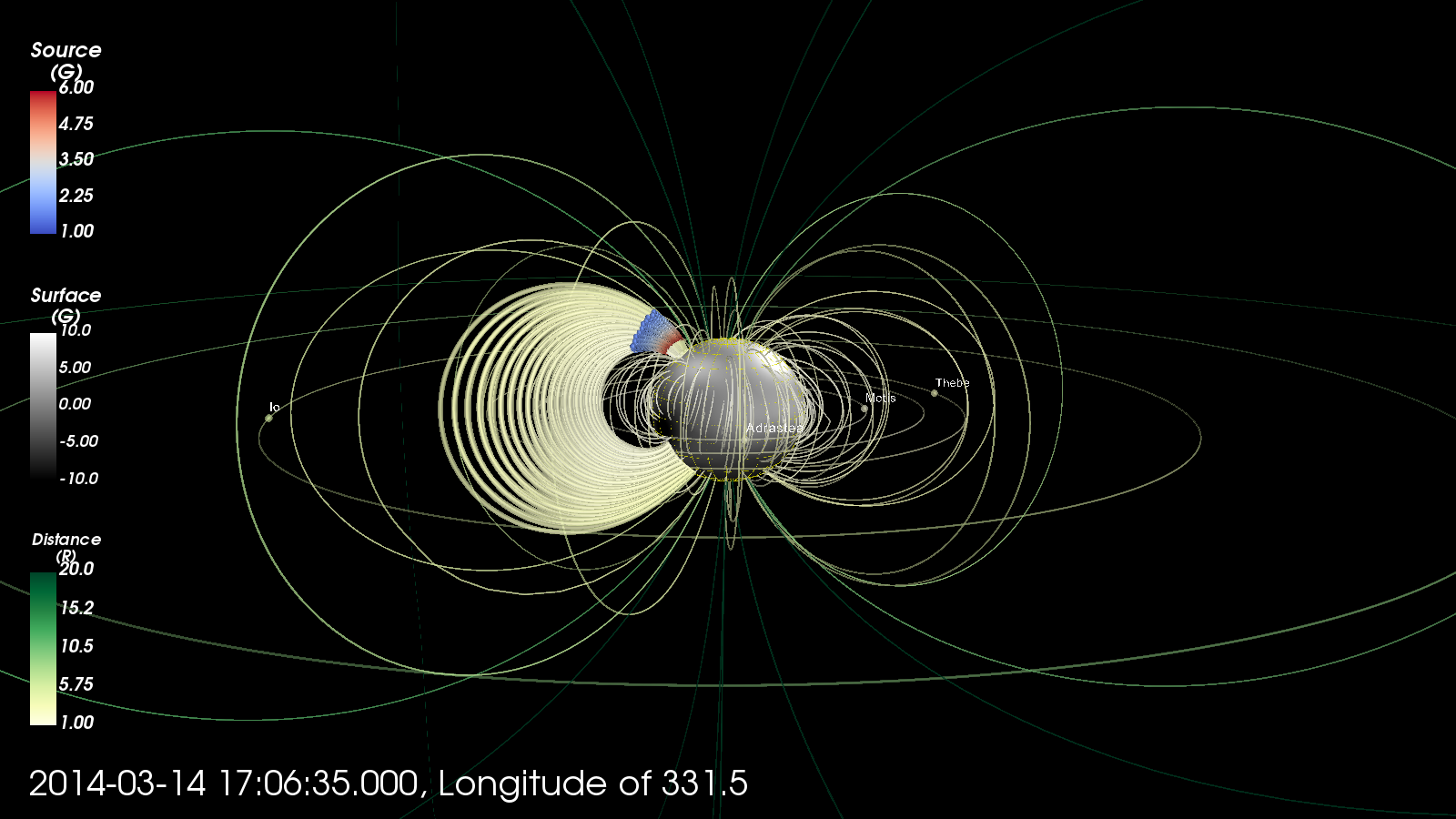}\\
\includegraphics[width=0.7\hsize]{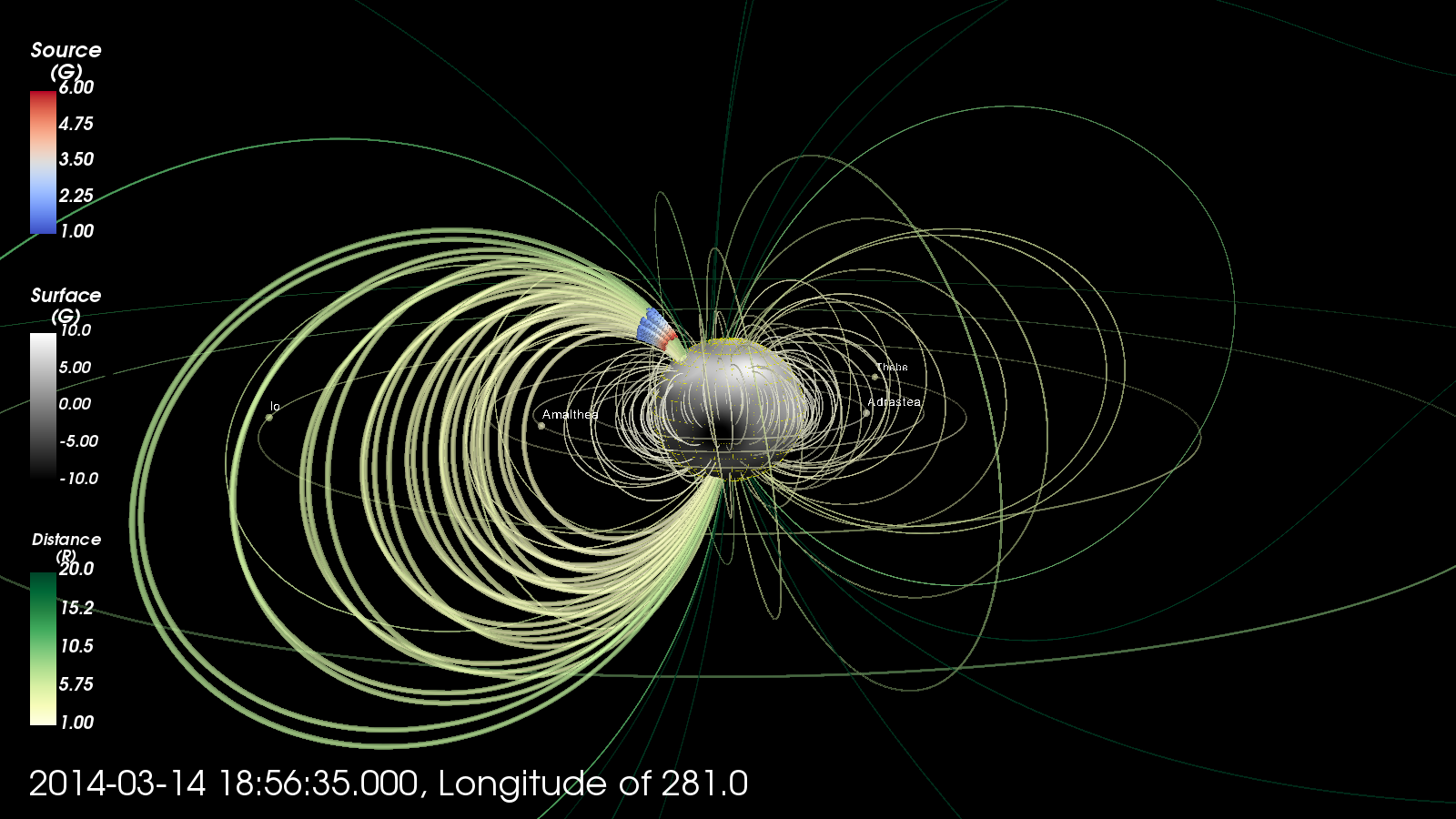}\\
\includegraphics[width=0.7\hsize]{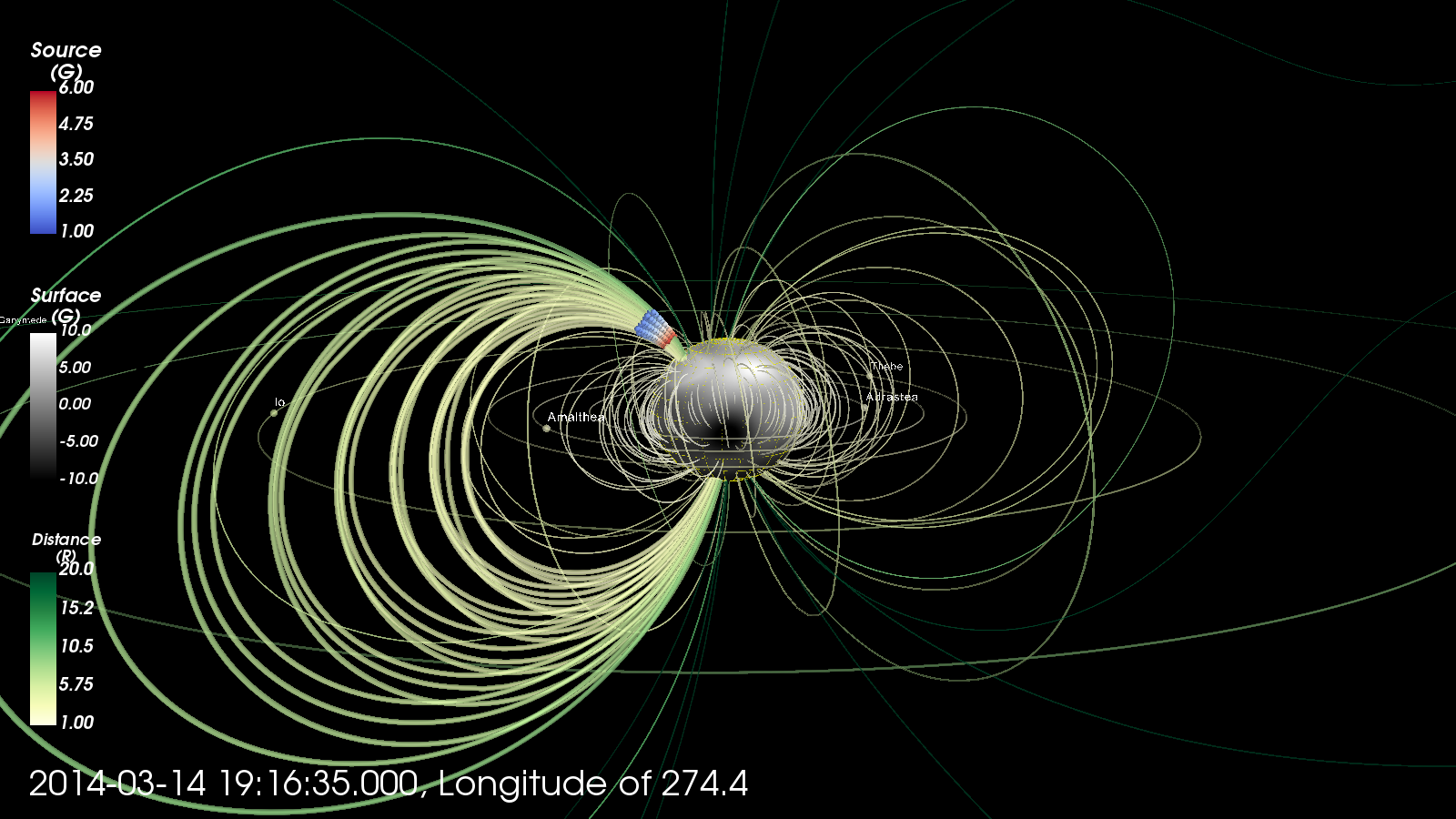}
	\caption{\small The upper to lower panels show the located DAM source based on the radio dynamic spectrum from the
Wind/WAVES, ST-A/WAVES and ST-B/WAVES, respectively.
The $1/15.4$ flattened surface of $B_r$ at one $\rm{R_J}$
is indicated by the gray-scaled ellipsoid at the center and is scaled by the gray bar on the left. Thin
curves show the background magnetic field lines with colors, scaled by the green bar on the left, denoting the
distances of the tops of these field lines away from the center of Jupiter. All the selected field lines
are thicker ones, and the located DAM source corresponding to the emission frequency from $5-16$ MHz
is marked by color-coded dots along these field lines. The color of the
DAM source indicates the local magnetic field strength, and is scaled by the color bar on the left too. The major moons,
including Io, Europa, Ganymede, Thebe, Amalthea, Adrastea and Metis, as well as their orbits are marked by small balls
and nearly round curves. The time and longitude indicated at the lower-left corner of each panel give the
time at Jupiter and the longitude of the disk center of Jupiter, respectively. These images are produced by using
Python with Mayavi~\citep{Ramachandran_Varoquaux_2011}. An animation could be found in the supplement.}\label{fig:source}
\end{center}
\end{figure*}

The located DAM source corresponding to the frequency range of $5-16$ MHz is shown in Figure~\ref{fig:source}.
From the upper to lower panels, there are the results based on the radio dynamic spectra
from the Wind, ST-A and ST-B spacecraft. It is found that
the radio source at a certain time is successfully confined in a relatively small region as indicated by the blue-red
colors. The red symbols correspond to the emissions at $16$ MHz. Their heights from the one-$\rm{R_J}$ surface
were about $0.1\rm{R_J}$ when Wind/WAVES
received the signals and increased to about $0.25\rm{R_J}$ when ST-A and B/WAVES received the signals.
The blue symbols correspond to the emissions at $5$ MHz with the height of about $0.65-0.75\rm{R_J}$ during the event.
All of these heights are well above the Jupiter's ionosphere.
Besides, the source field lines concentrate in a narrow angle in longitude, forming a fan-shaped sector. This sector
was moving ahead of Io as expected. A more detailed analysis is given below.

\section{Analysis of the Io-DAM source}\label{sec:analysis}

\begin{figure*}[tb]
\begin{center}
\includegraphics[width=0.595\hsize]{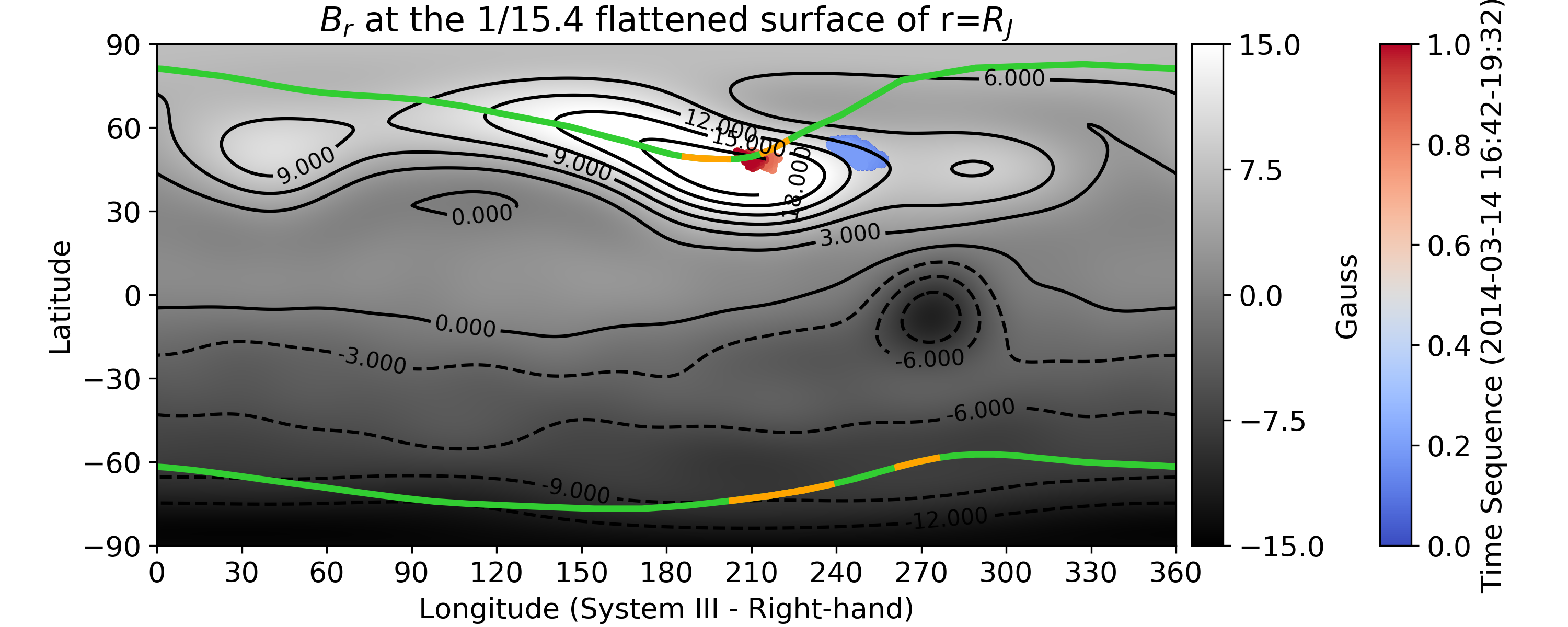}
\includegraphics[width=0.395\hsize]{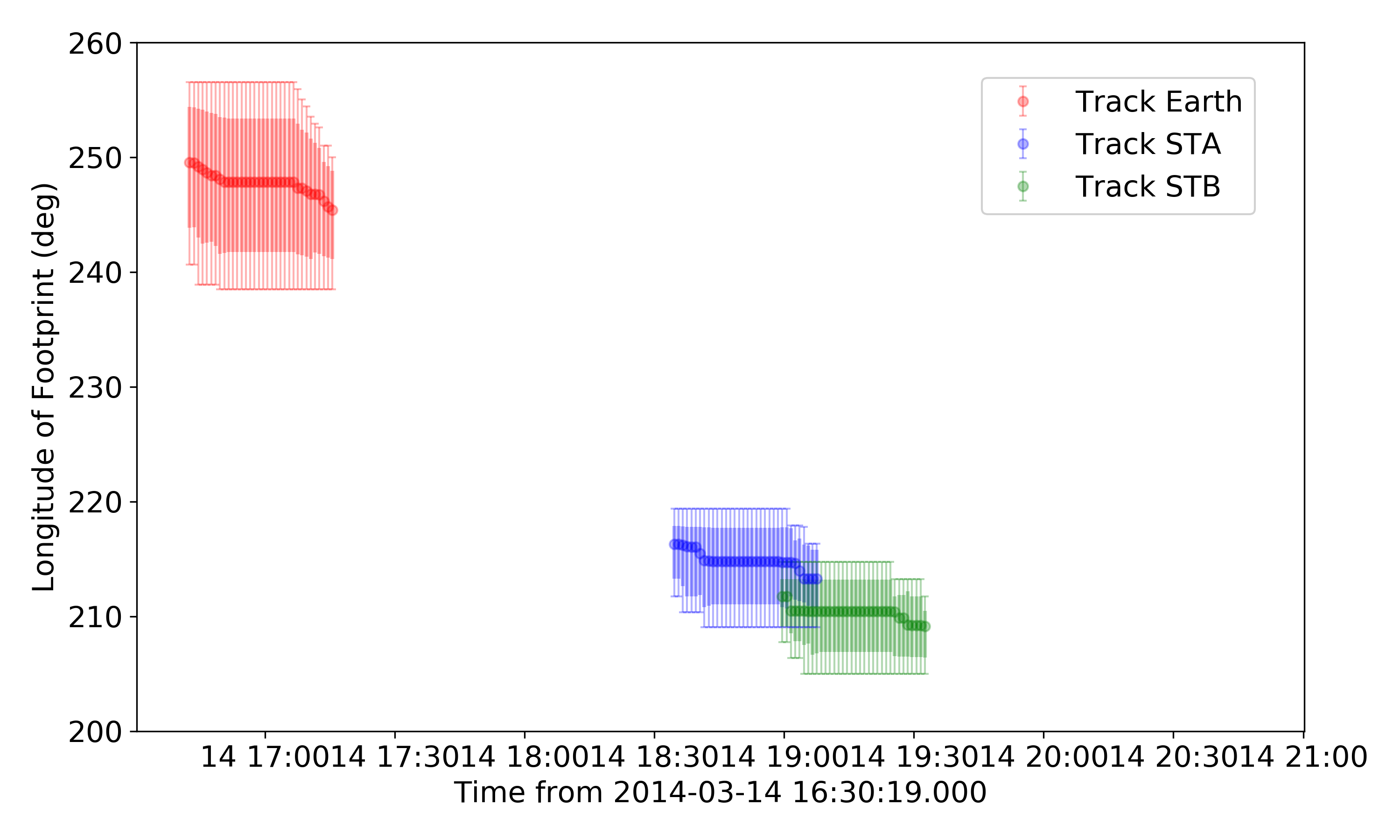}
\caption{The left panel shows contours of $B_r$ in Gauss at the $1/15.4$ flattened surface of one $\rm{R_J}$, which is
plotted in the right-handed System III coordinates. The color-coded symbols denote the footprints of selected
field lines where the DAM were emitted. The blue-to-red symbols correspond to the earlier to later times as indicated
by the color-bar on the right. Concretely, the blue symbols are obtained Based on Wind/WAVES observations, the orange
and red symbols based on ST-A and ST-B observations, respectively. The two green lines on the map denote the footprints
of Io according to the work by \citet{Bonfond_etal_2009}, and the orange segments mark the footprints when the
DAM observed by Wind, ST-A and ST-B. The right panel shows the longitudes of the footprints of these
selected field lines. The circles give the median values, the thick bars give the top and bottom 10th percentiles,
and the thin bars the minimum and maximum values.}\label{fig:surface}
\end{center}
\end{figure*}

The footprint of the identified radio source on the $1/15.4$ flattened surface at one $\rm{R_J}$ is displayed in
Figure~\ref{fig:surface}. Our method suggests that the source field lines were located near the edge of the region
that contains the strongest magnetic field in the northern hemisphere during 16:42 -- 17:15 UT (the left panel). Note that all the times hereafter
are the times at Jupiter if not specified. About two hours later, the footprint moved toward the strongest magnetic field region.
The change of the longitudes of these identified source field lines is shown clearer in the right panel, decreasing from about $248^\circ$ at Wind
to about $210^\circ$ at ST-B. Compared with the IFPs obtained from UV observations~\citep{Bonfond_etal_2009}, we may find
that the footprints of the source field lines located ahead of the IFPs (indicated by the orange line segments
in the left panel of Fig.\ref{fig:surface}) with a nearly constant lead angle of about $32^\circ$ in longitude.
It should be noted that the derived longitude only show changes during the first $10$ minutes and the last
$10$ minutes. This is because the DAM pattern on the radio dynamic spectra is an about $10$-minute-width drifting feature over 35 minutes;
during the middle 15 minutes, the inputs of our method are the same and therefore the derived parameters are the same.

The apparent rotation speed of the footprint relative to Jupiter magnetic field was about
$38^\circ/137$ min $\approx0.28^\circ/$min, or converting to the rotation speed in the inertial coordinates, $\Omega_{fp}\approx0.55\Omega_J$,
as same as the apparent longitudinal speed $\Omega_{IFP}$ on the northern hemisphere estimated in the last section.
This value is larger than $\Omega_{cc}\approx0.255\Omega_J$ derived from 2D correlation analysis above.
The reason is that the Jupiter's magnetic field
is not a pure dipole field with its axis aligned with Jupiter's rotational axis and the emission cone angles changed. To explain
$\Omega_{cc}<\Omega_{fp}$, the angle of main emissions should become smaller with time. As can be seen in
Figure~\ref{fig:ea}, the average values of the derived emission angles did change.
For instance, in the left panel, the average and maximum values of the emission angle both decreased by about $2^\circ$
during the 137 minutes.

\begin{figure*}[tb]
\begin{center}
\includegraphics[width=0.495\hsize]{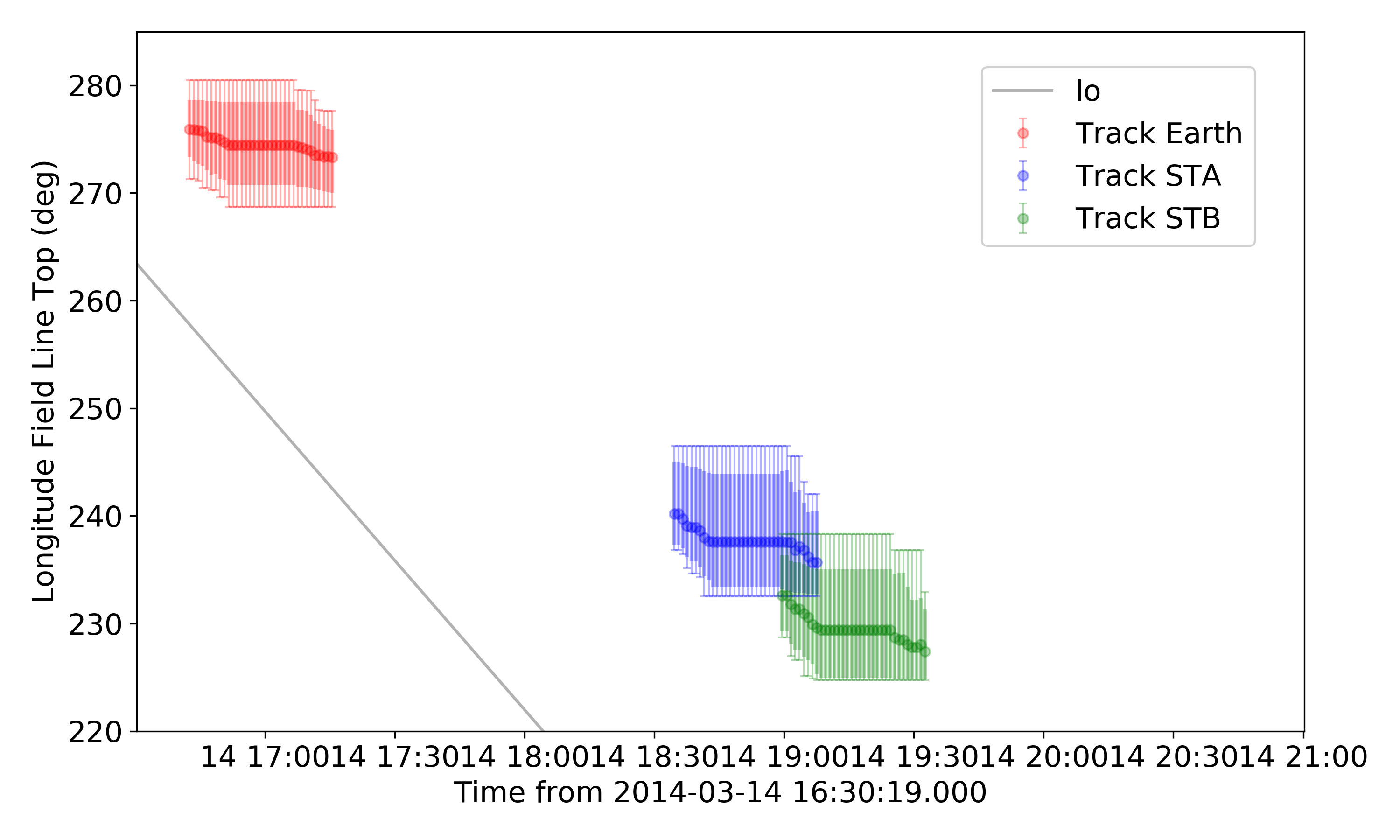}
\includegraphics[width=0.495\hsize]{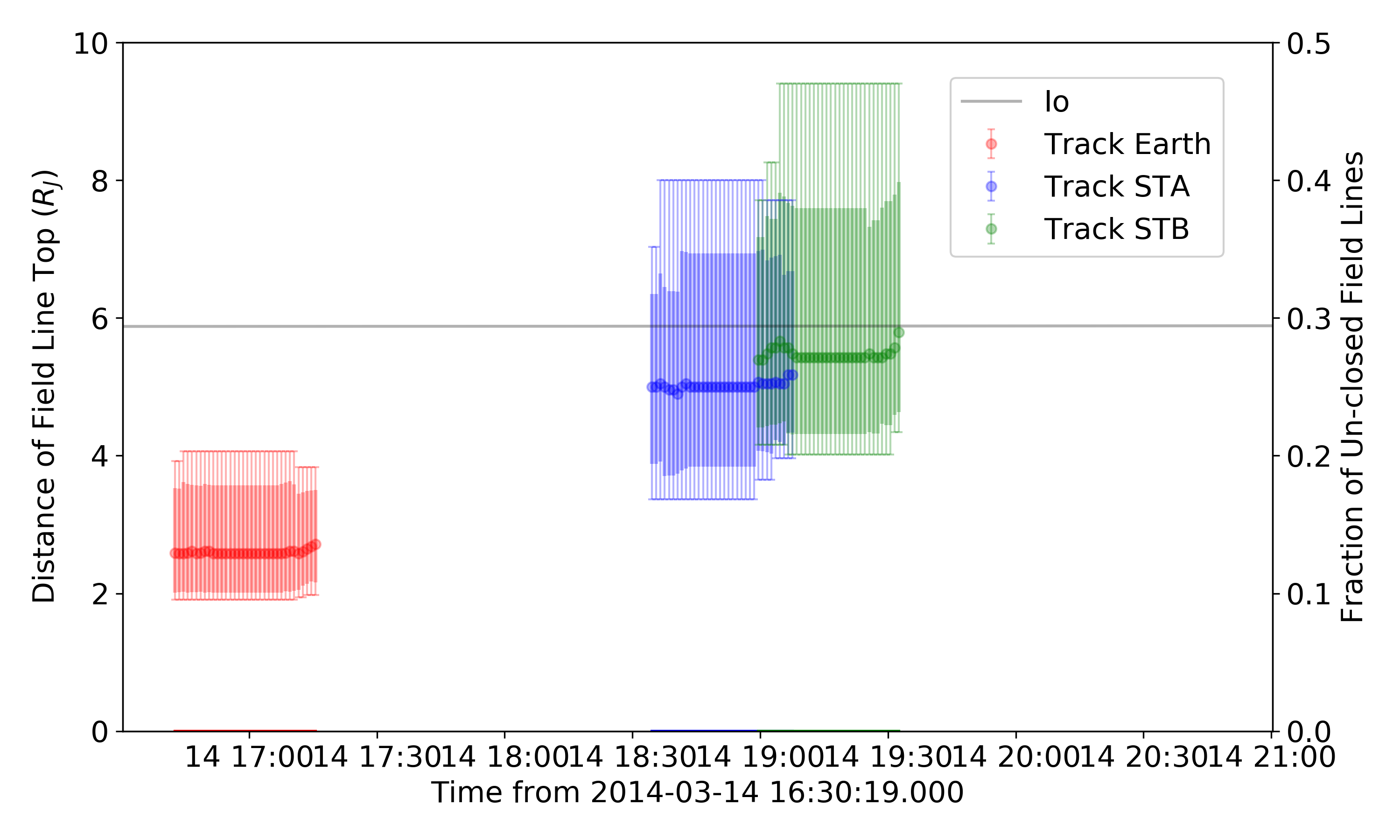}
	\caption{The left panel shows the longitudes of the tops of these
selected field lines. For comparison,
the gray line indicates the longitude of Io. The right panel shows the distances of the tops of the selected field lines.
In our procedure to calculate these values, all the open field lines or field lines with their tops beyond $50 \rm{R_J}$
are ignored. The fraction of the ignored field lines is indicated by the line scaled by the vertical axis on
the right. In this case, the fraction is zero. The circles
and bars in both panels have the same meaning of those in the right panel of Fig.\ref{fig:surface}.}\label{fig:rtop}
\end{center}
\end{figure*}

\begin{figure*}[tb]
\begin{center}
\includegraphics[width=0.495\hsize]{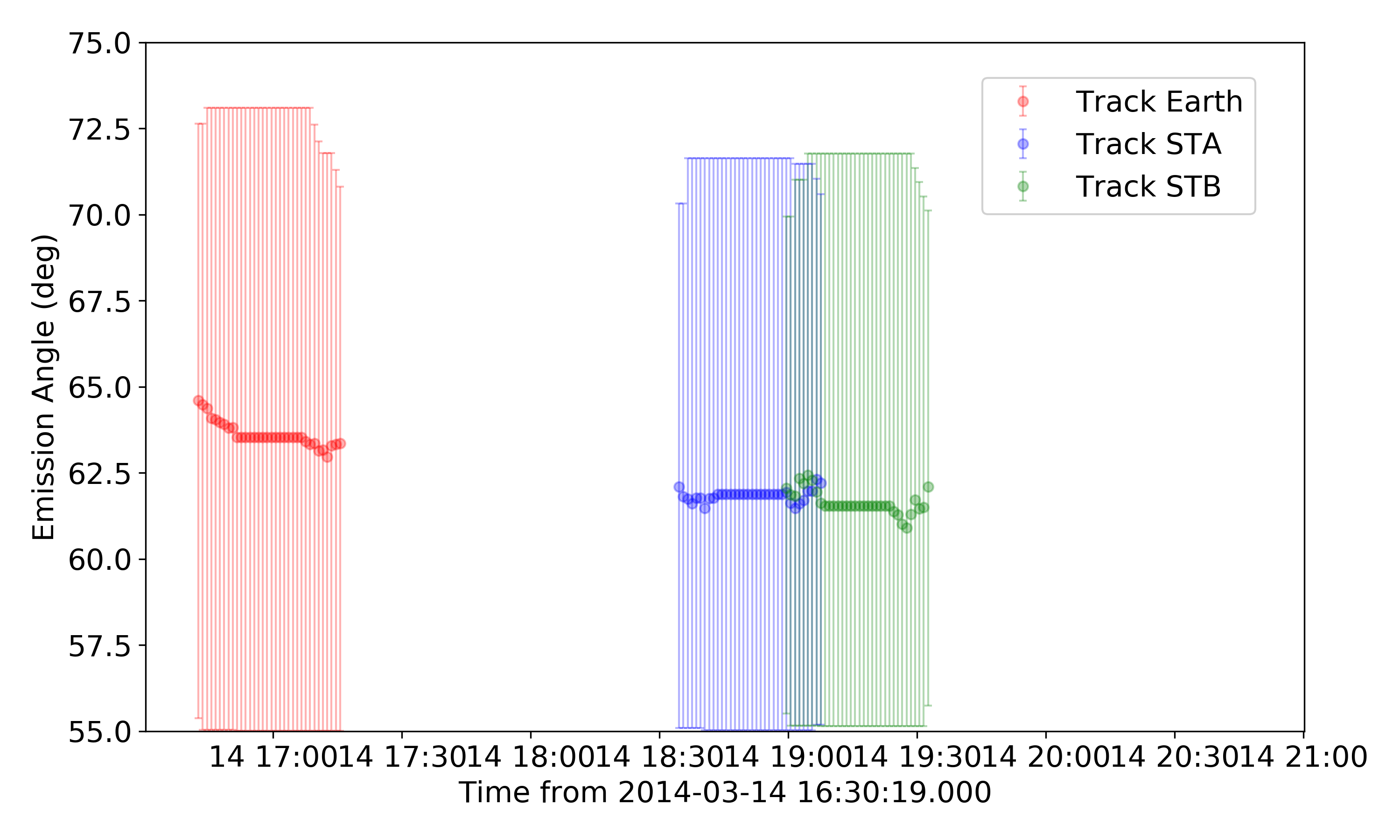}
\includegraphics[width=0.495\hsize]{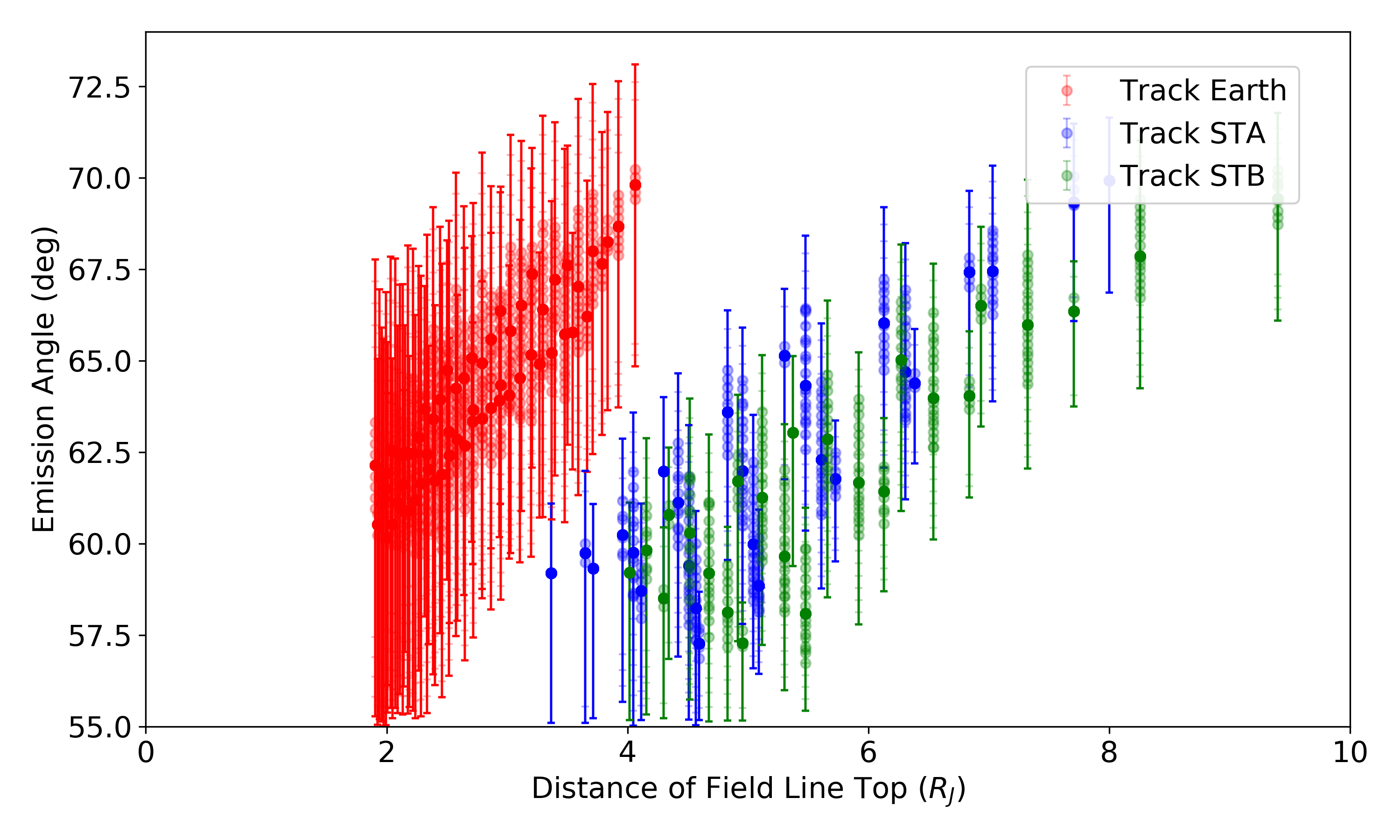}
	\caption{The left panel shows estimated emission cone angles during the DAM emissions.
The circles give the median values, and the bars give
the minimum and maximum values. The right panel shows the dependence of estimated emission cone angles on
the distances of the field line tops. Each circle with an error bar gives the median, minimum
and maximum values on a certain field line at a certain time.}\label{fig:ea}
\end{center}
\end{figure*}

The longitudes of the tops of these source field lines are shown in the left panel of Figure~\ref{fig:rtop}.
It is found that the longitudes of tops are different from those of footprints because of
the non-pure dipole field of Jupiter and the presence of the tilt angle between the magnetic axis and rotational axis as mentioned before.
These longitudes are compared to the longitude of Io to reveal possible relationship
between the source field lines and Io. When Wind/WAVES received the emissions, the longitude was about $24^\circ$ larger than that of Io.
And $137$ minutes later, i.e., when ST-B/WAVES received the emission, the difference increased to $42^\circ$, suggesting that the
apparent rotation speed of the source field line tops, $\Omega_{top}$, was about $0.45\Omega_J$, smaller than that of footprints but still
larger than $\Omega_{cc}$. It should be noted that the apparent rotation speed is not the rotation speed of any given field
lines, because the radio source originating from Io activities is not fixed on certain field lines. Moreover, $\Omega_{top}$ is not an
apparent speed at a certain distance, because the source field lines gradually shifted outward as shown in
the right panel of Figure~\ref{fig:rtop}.
The median value of the distances of the tops was about $2.5\rm{R_J}$
with the minimum of about $2\rm{R_J}$ and maximum of $4\rm{R_J}$ when Wind/WAVES
received the emission. About two hours later, the source field lines spread over a wider distance range from $3\rm{R_J}$
to nearly $10\rm{R_J}$ with most source field lines within the distance of $8\rm{R_J}$.

The emission cone angles derived by using Equation~\ref{eq:ea} through the fitting
procedure in our method are displayed in Figure~\ref{fig:ea}. A pattern is clearly revealed in the right panel that
the emission angle tends to be smaller on the inner magnetic field lines than that on the outer magnetic field lines.
The derived emission angles fall well
within the preset range of $50^\circ-90^\circ$ in the method. Besides, as mentioned before, the averaged emission angle decreased with
time during the event as shown in the left panel. This is a cause of that the apparent rotation speed of the source, i.e., $\Omega_{cc}$,
was slower than that of source field lines.

The energy (or the speed) of energetic electrons exciting the DAM emissions is the fitting parameter of Equation~\ref{eq:ea}
and plotted in Figure~\ref{fig:ele}. The preset electron energy is $0.2$ keV, corresponding to the speed of $0.05c$. The
derived electron energy and speed are well above the limits, suggesting a successful fitting. According to the left panel,
it is suggested that when the emission cone swept through Wind spacecraft, the energetic electrons confined in the source
field lines had an energy in the range of about $7.5-15.5$ keV, and 137 minutes later, when it swept through ST-B, the energy
of the associated electrons spread into a slightly wider range of about $9-24$ keV.
The electron energy falls in the similar range fitted by \citet{Hess_etal_2007, Hess_etal_2010} using the Io-DAM arcs,
but are slightly larger than the energy of electrons emitting the millisecond bursts.
The dependence of the electron energy
on the distance of the field line top given in the right panel shows the change more clearly.
The electrons emitting the DAM observed by ST-A and ST-B were apparently more energized than those observed by Wind,
with energy increasing from below $\sim15$ keV to upto $24$ keV.
The electrons on the outer magnetic field lines typically had a lower energy than those on the inner magnetic field lines.
Moreover, according to Equation~\ref{eq:ea}, the emission angle decreases with the increase of electron energy. Thus,
the emission angle on the inner magnetic filed lines was smaller than that on the outer magnetic field lines as shown
in the right panel of Figure~\ref{fig:ea}.

\section{Summary and Discussion}~\label{sec:summary}

In this paper, we present a method to locate the source of DAM emissions from Jovian magnetosphere following
the work by~\citet{Hess_etal_2008, Hess_etal_2010}.
The method only uses the time and frequency of the DAM drift pattern in the radio dynamic spectrum
recorded by a spacecraft, and allows us to derive the information of the source location, source field lines,
emission cone angle and the energy of associated electrons. If there are multiple spacecraft receiving the DAM emission from
different perspectives, the evolution of the DAM source can be revealed.

By applying this method to an Io-DAM observed on 2014 March 14, we locate the source field
lines rooting in a relatively small region in Jupiter's northern ionosphere close to where Io is
magnetically mapped, just as expected. Considering the change of emission angles,
we find that the rotation speed of the emission source, i.e., the rotation speed of the footprints of
identified source field lines, matches that estimated from 2D correlation analysis.
The derived emission cone angles and electron energies are well within the preset ranges.

\begin{figure*}[tb]
\begin{center}
\includegraphics[width=0.495\hsize]{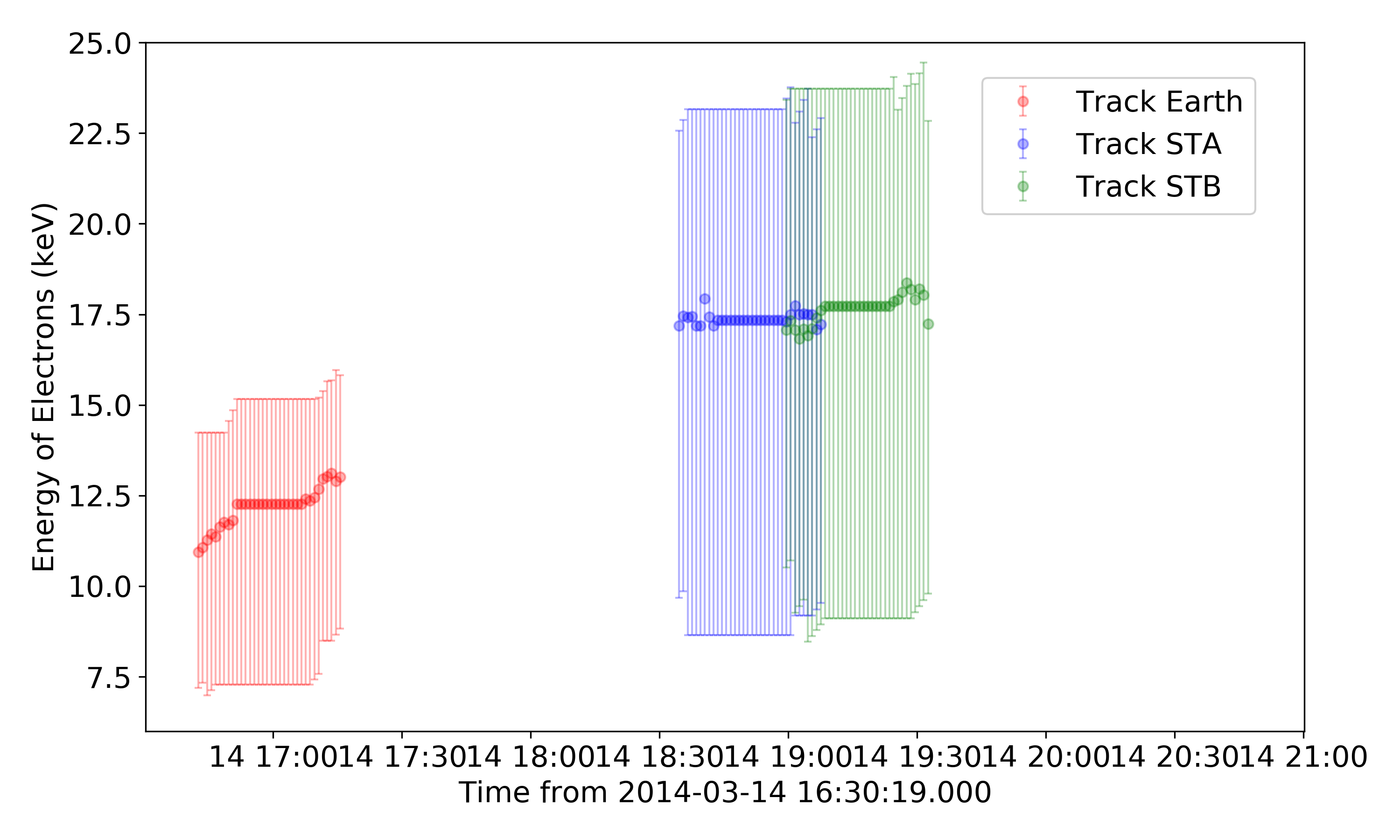}
\includegraphics[width=0.495\hsize]{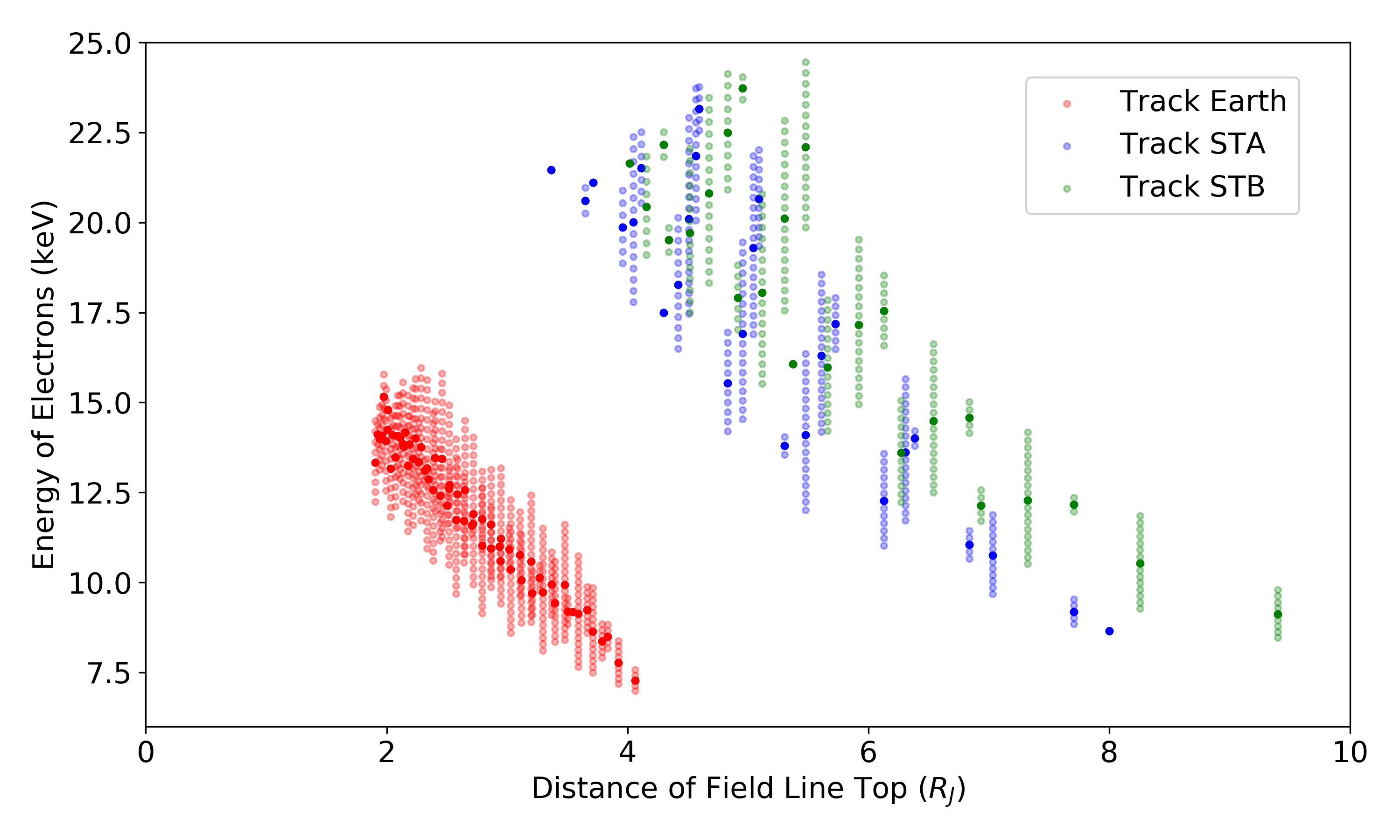}
	\caption{The left panel shows the estimated electron energy. The circles with error bars indicate the
median, minimum and maximum values. The right panel shows the dependence of the electron energy on the
distances of the field line tops.}\label{fig:ele}
\end{center}
\end{figure*}

\begin{table*}[t]
\footnotesize
\caption{Some characteristic parameters}\label{tb:par}
\begin{tabular}{c|cc|ccc|ccc}
\hline
$\Omega_{cc}$ & $\Omega_{Io}$ & $\Omega_{IFP}$ & LA$^a$ & $\Omega_{fp}$ & $\Omega_{top}$ & Distance/$R_J$ & EA$^b$/$^\circ$ & Energy/keV \\
\hline
$0.255\Omega_J$ & $0.23\Omega_J$ & $0.55\Omega_J$ & $32^\circ$ & $0.55\Omega_J$  & $0.45\Omega_J$ & $\sim\left(2.5^{+1.5}_{-0.5} - 5.5^{+2}_{-1.5}\right)$ & $\sim\left(63.5^{+9.5}_{-8.5} - 61.5^{+10}_{-6.5}\right)$ & $\sim\left(12.5^{+3}_{-5} - 18^{+6}_{-9}\right)$ \\
\hline
\end{tabular}
$^a$ Lead angle of the footprints of the field lines.\\
$^b$ Emission cone angle.
\end{table*}

Further integrating the results presented in Section~\ref{sec:analysis}, we may outline the following picture
for this Io-DAM event. Volcanic activities on Io released neutral particles into space, which then rapidly
got ionized becoming plasma. Through additional acceleration and transport processes and
their interaction with Jupiter's magnetosphere, some of the plasma may be energized producing energetic electrons.
These electrons were trapped by magnetic field lines connecting to Io, moved down toward the surface of Jupiter
and excited DAM emissions through CM instability. Some electrons were precipitated into ionosphere,
and some were bounced in the field lines and gradually diffused depending on their energies and pitch angles.
During this process, the electrons were somehow further energized and spread over in a wider region. The
diffusive motion of these electrons led to that the DAM source was not fixed on initial
field lines, but propagated with a pace very close to Io's footprint. The apparent speeds of the footprints and tops of the DAM source field lines were about
$0.55\Omega_J$ and $0.45\Omega_J$, respectively. The obtained footprint speed is almost the same as that
of the IFP on the northern hemisphere with a nearly constant lead angle of about $32^\circ$,
suggesting a strong connection with Io. All the speeds were larger than the apparent rotation speed of the
DAM source derived from 2D cross-correlation analysis, which was about $0.255\Omega_J$, mainly due to
the decrease of emission cone angle. All characteristic parameters obtained in this study are
summarized in Table~\ref{tb:par}. This picture is consistent with our current understanding of Io-DAM
emissions~\citep[e.g.,][]{Zarka_1998, Hess_etal_2008}. All of the above results suggest that our method is valid.

Moreover, it is found that the electrons were more energized when the footprints of the DAM source filed lines
were located in a stronger field region as can be seen from Figure~\ref{fig:surface} and \ref{fig:ele}. \citet{Hess_etal_2010}
found a dependence of the electron energy on Io's longitude. We consider this may be due to the variation of the magnetic
field intensity of the flux tubes connecting with Io. The larger the magnetic field intensity is, the stronger can be
the interaction between Io and Jupiter, and the electrons would likely to be more energized. However, this may also be
a result from the assumption used in the model. In Equation~\ref{eq:ea}, the emission angle is a function
of the combination of the electron velocity and the magnetic field strength at the top of ionosphere. For this event,
the change of the average value of the derived emission angle was small. One can therefore expect that the electron energy
derived from Wind observation will be lower than that derived from ST-B observation, since the DAM received by Wind came
from a source region with a weaker magnetic field than that by ST-B.
Further analysis based on the infrared and UV observations may be helpful to clarify this.

According to the Alfv\'en wing model of Io-Jupiter magnetic field interaction~\citep{Hill_etal_1983},
the source field lines not corotating with Jupiter but moving slightly faster than Io is
attributed to the presence of the Io's ionosphere, of which the electrical conductance, $\Sigma_{I}$,
including the Pedersen and Hall conductance, is significant and
can slow down the Jupiter's field lines passing through it. Previous studies
suggested that the conductance is about two orders higher than the Alfv\'en conductance, $\Sigma_{A}$,
outside the Io's ionosphere~\citep[e.g.,][]{Saur_etal_1999, Kivelson_etal_2004}.
For such a high conductance, the field lines in the vicinity of Io would be dragged to almost corotate
together with Io for a while. When the field lines are dragged by Io, Alfv\'en wings will form and
result in the longitude difference between the source field lines and Io, which is called Mach angle and
typically less than $10^\circ$ (refer to Fig.10.3 in~\citet{Hill_etal_1983}).
For this event, our method shows that the tops of source field lines located ahead of Io with a lead
angle from about $24^\circ$ at Wind to about $42^\circ$ at ST-B.
This is different from that for the footprints of the source field lines, of which the lead angle is
about $32^\circ$ without significant change during the event.

Besides, at the earlier time, i.e., when the DAM emission lighted Wind spacecraft, the source field lines
were within the Io orbit. We are not sure if this is true or resulted from some uncertainties in our method.
If it is true, the energetic electrons should be injected and/or quickly transported into inner magnetic field lines
through some process. A possible scenario is as follows. Io ejected plasmas into magnetosphere. These plasmas were trapped
by and corotated with Jupiter's field lines, and stretched these field lines outward due to the centrifugal
force. The stretched field lines began to reconnect and convect inward at a certain time and consequently generated
energetic electrons. These electrons were further transported into the inner magnetosphere within the Io orbit through
cross-field diffusion. However, since only the static magnetospheric
field model is used in our method, we are unable to distinguish any detailed dynamic processes.

\begin{acknowledgments}
We acknowledge the use of the data from the radio instruments on board Wind, STEREO-A and B spacecraft.
We thank the anonymous referees for many valuable comments.
We are also grateful to Hao Cao from Harvard University for valuable discussions.
The work is support by the Strategic Priority Program of the Chinese Academy of Sciences (Grant Nos. XDB41000000,
XDA15017300) and the NSFC (Grant No.41842037 and 41574167).
V.K. acknowledges support by an appointment to the NASA postdoctoral program at the NASA Goddard Space
Flight Center administered by Universities Space Research Association under contract with NASA and the
Czech Science Foundation grant 17-06818Y.
\end{acknowledgments}

\bibliographystyle{agu}
\bibliography{../../ahareference}

\end{article}
\end{document}